\documentclass[%
 reprint,
 amsmath,amssymb,
 aps,
prb,
floatfix,
]{revtex4-2}

\usepackage{rotating}
\usepackage{graphicx}
\usepackage{dcolumn}
\usepackage{bm}
\usepackage{amsmath}
\usepackage{amssymb}
\usepackage{subfigure}  
\usepackage{hyperref}
\usepackage{nicefrac}
\usepackage{hyperref}
\hypersetup{colorlinks=true, breaklinks=true, citecolor=blue, linkcolor=blue, menucolor=blue, urlcolor=blue}
\usepackage{ulem}
\usepackage[dvipsnames]{xcolor}

\newcommand{\threerectcell}{$(3\times \sqrt 3 )_\textrm{rect}$}
\newcommand{\rectcell}{$\left(5 \times \sqrt{3}\right)_{\text{rect}}$}

\newcommand{\cutwotetwo}{Cu$_2$Te$_2$}

\begin{document}

\preprint{APS/123-QED}

\title{New submonolayer copper telluride phase on Cu(111) - ad-chain and trough formation}

\author{Tilman Ki{\ss}linger}
\author{M. Alexander Schneider}%
\email{alexander.schneider@fau.de}
\author{Lutz Hammer}%
\email{lutz,hammer@fau.de}
\affiliation{%
 Solid State Physics, Friedrich-Alexander-Universität Erlangen-Nürnberg, Staudtstraße 7, 91058 Erlangen, Germany
}%

\date{\today}

\begin{abstract}

We present a so far undetected submonolayer phase of copper telluride on Cu(111) with \rectcell\ periodicity and coverage of 0.40 ML Tellurium (Te), which can be grown with perfect long-range order. 
It is structurally characterized by a combination of quantitative low-energy electron diffraction (LEED-IV), scanning tunneling microscopy (STM), and density functional theory (DFT). 
We find that Te induces the formation of four atom wide linear troughs within the Cu(111) surface filled up by two Te atoms per unit cell. Additionally, the interspace between the troughs is decorated by \cutwotetwo\ ad-chains sitting at hcp sites. All Te atoms exhibit the same local sixfold coordination: They occupy threefold hollow sites of the substrate and are one-sided attached to another three Cu atoms. 
The presented structural model is verified by a LEED-IV analysis with Pendry R-factor of R = 0.174 and quantitative agreement of structural parameters with DFT predictions. It also has by far the lowest energy of all models tested by DFT and simulated STM images agree perfectly with experiment. 
It turns out that the new \rectcell\ phase is the most dense surface telluride phase possible on Cu(111) before bulk-like copper telluride starts to growth.
\end{abstract}

\maketitle


\section{\label{sec:intro} Introduction}

Climate change and its future consequences are currently the biggest challenges to be tackled by humankind and science \cite{Tollefson2018,ONeill2017}. Considering a constant or even rising demand of power, the generation of environment-friendly energy requires solutions like e.g. highly efficient photovoltaic modules or hydrogen based fuel cells \cite{Bogdanov2019,Gernaat2021}.

Generally, transition-metal dichalcogenides (TMDs) can be used as materials in solar cells and for energy conversion from hydrogen \cite{Singh2017,Li2018,Iqbal2019}.
Copper tellurides in particular are e.g. of great importance as back-contacts in CdTe/CdS photovoltaic cells  due to their potential to increase the electrical conductivity at the absorber-contact interface. The efficiency of the interface is furthermore strongly influenced by the amount of Cu ions diffusing into the CdTe/CdS bulk structure leading to an increased interest in the structure and chemical stability of such materials \cite{Teeter2007,Perrenoud2013,Xia2014,SalGam2018,Du2019}. 

On the other hand TMDs are promising candidates to catalyze the hydrogen-evolution reaction \cite{Kong2013, Wang2017, Lee2018} and thus may play a major role in the decarbonization of industry \cite{Ibers2009, Chivers2015a}.
With the same objective, telluride based thermoelectrica exhibiting large figures of merit like Bi$_2$Te$_3$ \cite{Witting2019} but also copper telluride must be mentioned \cite{Mansour1986,Nethravathi2013,He2015}. 

As known for long, the precise knowledge on the nature and atomic arrangement of so-called active sites is important for any catalytic reaction \cite{Somorjai1977}. Based on that structural information it might be possible to develop models to measure and predict the reactivity and selectivity of different materials at the atomic level \cite{Ertl2008, Hwang2019}.

For the particular system Te on Cu(111) there has been a long-standing debate about the structure and coverage of the first long-range ordered phase with \mbox{$(2\sqrt 3\times 2\sqrt 3)R30^\circ$} LEED pattern  that is observed upon Te adsorption \cite{Andersson1968, Comin1982, King2012, Lahti2014}. This issue could be resolved recently by an investigation of our group \cite{Kisslinger2020} to host 1/3\,ML of Te atoms per unit cell consisting of \cutwotetwo\ chains at hcp-positions of an unreconstructed Cu(111) substrate. Taking this as a gauge to quantify the coverage of a subsequently observed $\left(\sqrt{3} \times \sqrt{3}\right)\textrm{R30}^\circ$ phase, we end up at an approximate Te content of 2/3\,ML \cite{King2012} or 4/3\,ML \cite{Andersson1968, Tong2020}. For this phase the authors of Ref.\,\cite{King2012} propose the growth of an unusual Cu$_3$Te$_2$ bulk alloy, whereas in \cite{Andersson1968} and \cite{Tong2020} a substitutional surface alloy was predicted, which, however, is not compatible with the now re-calibrated coverage. In total, there are no structural investigations in the coverage regime between 1/3\,ML and 2/3\,ML at all and the crystallographic structure of the phases observed for $\Theta \geq$ 2/3 ML is still under question.

Here, we report on a so far unknown submonolayer phase of Te on Cu(111), which evolves at a coverage of 0.40\,ML Te. We investigate in detail the formation of this phase, its stability and unequivocally reveal it crystallographic structure by a combination of quantitative LEED, room-temperature STM, and DFT.

\section{\label{sec:exp} Experimental Details}

All experiments were performed in an UHV-apparatus that is divided into two independently pumped chambers. The first one hosts all equipment for sample preparation as well as a 3-grid LEED optics and operates in the low $10^{-10}\,$mbar regime. The manipulator allows for rapid heating of the sample by electron bombardment from the rear and cooling to a minimum of $90~$K within minutes by direct contact to a liquid-nitrogen reservoir.
The second part has a base pressure of $2 \cdot 10^{-11}\,\textrm{mbar}$ and contains a beetle-type room-temperature STM with tungsten tip and the bias voltage applied to the sample. Prior to Te deposition, the Cu(111) substrate was cleaned by several cycles of sputtering (Ne$^+$, 2\,keV) and oxygen-annealing at 870\,K until it was found to be clean in both, LEED and STM. High-purity Te (99.999 $\%$, lump, Alfa Aesar) was deposited from an electron-beam evaporator at a rate of $\approx 0.1~$ML/min. The coverage could be precisely controlled via calibration against the \threerectcell\ structure that evolves on Cu(111) for a \nicefrac{1}{3}~ML Te content \cite{Kisslinger2020}. Here, already deviations as small as 0.01 ML from the nominal coverage lead to the occurrence of easily detectable characteristic satellite spots in LEED.

\subsection{\label{sec:leed_data} LEED data acquisition}

LEED measurements were performed at normal incidence of the electron beam (within 0.1$^\circ$ accuracy). In order to reduce the influence of the thermal diffuse background the freshly prepared sample was cooled to $100~$K prior to any data acquisition.
A stack of LEED-images (so-called LEED-video) was recorded by a CCD camera in the energy interval between 50 and 500\,eV in steps of 0.5 eV and stored on a hard-drive for later off-line evaluation with the new \textsc{ViPErLEED} package \cite{Riva2021}. Inhomogeneities in the detection device (LEED optics and camera) as well as the energy dependence of the primary beam current were removed by normalizing the recorded and dark-current corrected intensities to a so-called flat-field video. 
For that the very same energy scan was recorded for the electron beam being directed at an amorphous ring made out of the substrate's material surrounding the sample. 
From this final stack of frames, intensity vs. energy curves (so-called IV-spectra) were recorded by a fully automated spot tracking system with sub-pixel accuracy \cite{Riva2021}. Remaining noise was subsequently removed by smoothing the resulting spectra four times with a Savitzky-Golay filter (4th order, 21 points). Artefacts eventually remaining from background subtraction were removed by cutting off the affected low or high energy range of the spectrum.
This led to a total set of 79 symmetrically inequivalent beams with a cumulated energy range of $\Delta E \approx $17.5\,keV, whereby most of the data base (15.4\,keV) consisted of structurally most sensitive fractional order beams (72 beams).

\subsection{\label{sec:leed_calc} LEED-IV calculations}

Full-dynamic LEED intensity calculations were performed using the extended capabilities of the \mbox{\textsc{ViPErLEED}} package \cite{Riva2021}, which manages a modified \mbox{\textsc{TensErLEED}} code \cite{Blum2001}.
The slightly convergent electron beam of the experiment was modeled by a small off-normal angle of incidence (adjusted to 0.35$^\circ$ within the LEED-fit) and averaging over various directions. Phase shifts were used up to angular momentum of $l_{max} = $14 and calculated together with the real-part of the inner potential $V_{0r}$ by Rundgren's program {\small EEASiSSS} \cite{Rundgren2003}, which is also implemented in the \mbox{\textsc{ViPErLEED}} package. The optical potential was taken constant and fitted to $V_{0i} =$ 3.8\,eV during the analysis. The lattice parameter for Cu at 90\,K was taken from Ref.\,\cite{LattConstCu} and set to 2.550\,\AA. The bulk vibrational amplitude was set to 0.0875\,\AA, according to a Debye temperature of \mbox{$\Theta_{\text{D}}=$ 345\,K} (as an average over a variety of published values). 
The correspondence between model intensities and the experimental IV-spectra was quantified using Pendry's R-factor \cite{Pendry1980}. Parameter optimization was performed by a search algorithm \cite{Kottcke1997} within the TensorLEED approach \cite{Rous1986, Heinz1995} that is included in the \textsc{TensErLEED} package.

We optimized atomic coordinates up to the $3^{rd}$ substrate layer since DFT calculations predicted for all investigated models negligible deviations from bulk positions below. 
For the initial model discrimination the optimization was restricted to vertical parameters as well as lateral parameters of the Te and adlayer Cu atoms to a precision of 0.01\,\AA. 

For the final best fit we included all vertical and lateral parameters within the investigated surface slab as well as vibrational amplitudes of all undercoordinated atoms with a final precision of 0.002\,\AA\ each. In total, $N = 56$ parameters (thereof 48 geometrical and 5 vibrational) were varied in the analysis. This leads to a redundancy factor $\rho = 4NV_{0i}/\Delta E = 20.3$  of the LEED analysis, i.e. we had in principle twenty times more data than needed at minimum for the fit because of the enormous experimental data base.

An estimation of the statistical error of the fit parameters was obtained by variation of single parameters out of their best-fit value up to the point where the variance level of the R value ($\text{var(R)} = R \cdot \sqrt{8V_\text{0i}/\Delta E}$) is intersected. In order to reduce the computational effort we abstained from angular averaging and calculated these ``error curves'' for one azimuthal orientation only, so that the minima of the curves are not exactly at the bestfit R-factor value but by about 0.003 higher, which had no measurable effect on the size of the error margins. 

\begin{figure*}[htbp]
	\centering
	\includegraphics[width=\textwidth]{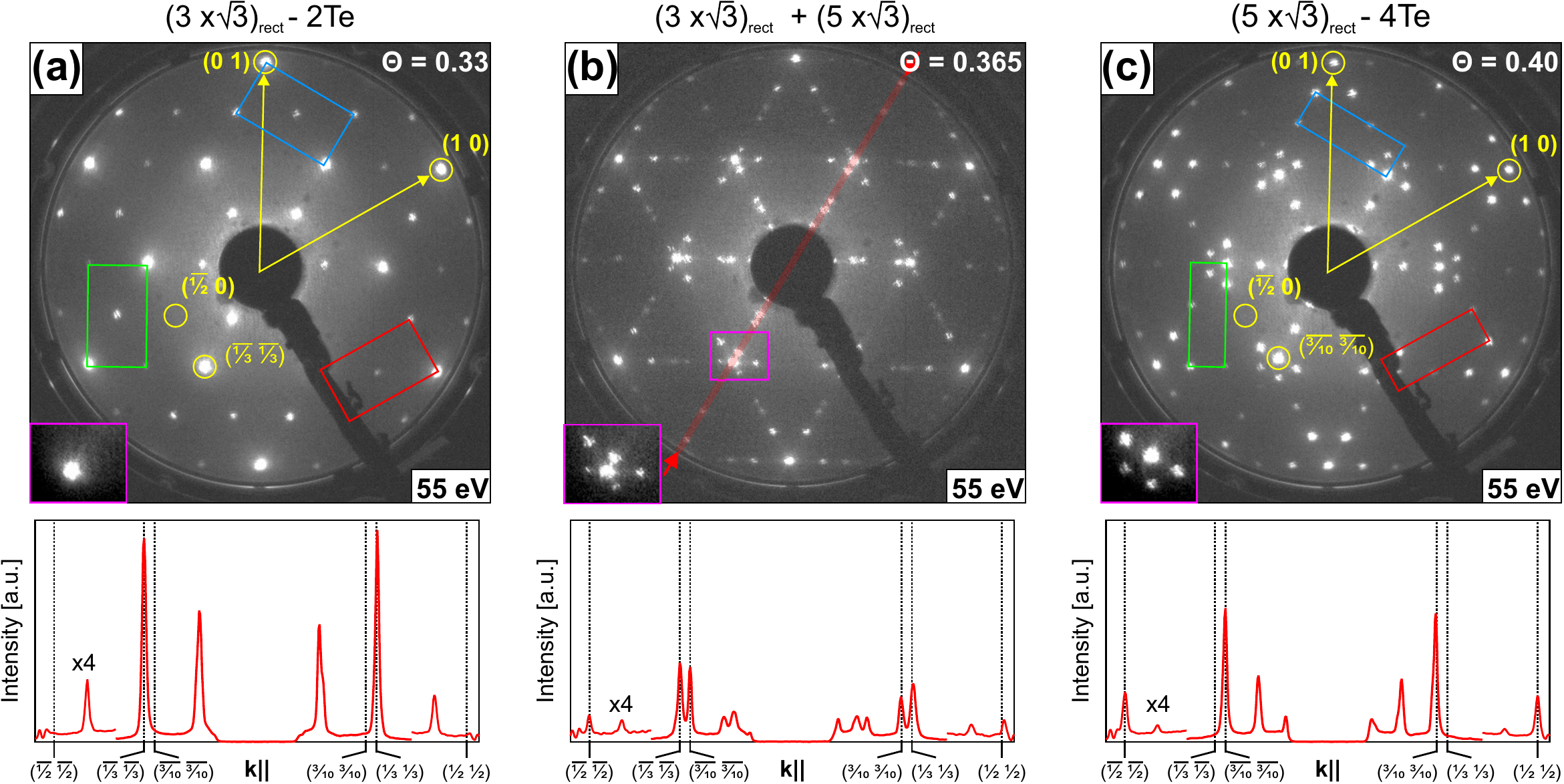}
	\caption{LEED pattern (55\,eV; top) and corresponding intensity profiles taken in [1 1] k-space direction (along the red line shown in (b); bottom) for various Te coverages after subsequent annealing to 750\,K. (a) \threerectcell\ phase at $\Theta = 0.33$\,ML in three domains (unit cells marked in red, blue, green). (b) Coexistence of \threerectcell\ and the new \rectcell\ phase for $\Theta = 0.365$\,ML. (c) Fully developed \rectcell\ phase at $\Theta = 0.40$\,ML in three domains. 
		For both phases we notice vanishing half-order beams along the main k-space axis (see e.g. the $(\overline{\nicefrac{1}{2}}~0)$ spot, whose position is marked by a yellow circle) indicating a glide plane within the structures.	For further details see the main text.
	}
	\label{Fig1}
\end{figure*}

\subsection{\label{sec:dft_calc} Density Functional Theory Calculations}

DFT calculations were performed using the Vienna Ab-initio Simulation Package (VASP) \cite{vasp3} employing the PBE-PAW general gradient approximation \cite{PBE} with energy cutoff of 385\,eV using a ($3\times 10\times 1$) $\Gamma$-centered k-point mesh.
Spin-orbit coupling was not taken into account.
A \rectcell\ slab consisting of 6 layers of Cu(111) and the Te containing layer(s) on the top side with the three bottom layers kept fixed at bulk positions. 
The cell size in slab normal direction was chosen equivalent to 15 Cu(111) layers leading to a separation of repeated slabs by at least 15\,\AA\ of vacuum. 
The structures were relaxed until forces were smaller than 0.01\,eV/\AA.
STM images were simulated using the Tersoff-Hamann approximation \cite{Tersoff85}.

The difference in the phase formation energy $\Delta E_f$ between the various investigated models and the LEED bestfit model, which was also the energetically most favorable structure, was calculated by 
\begin{equation}
\label{eq:formationE}
\Delta E_f = E_{\sigma} - n\cdot E_{Cu} - E_{bf}  \text{.}
\end{equation}

Hereby, $E_{\sigma}$ and $E_{bf}$ are the total energies of the phase $\sigma$ and the LEED bestfit structure and $n \cdot E_{Cu}$ is the total (bulk) energy of the $n$ surplus Cu atoms within the phase $\sigma$ compared to the bestfit structure ($n$ may be positive or negative). The Cu bulk energy  was determined as $E_{Cu} = -3.733 \pm 0.004$\,eV from total energy calculations of Cu(111) slabs with different thicknesses.\\

\section{\label{sec:res} Experimental Characterization of the New Phase}

\subsection{LEED investigations}\label{sec:res:leed}

The top row of Fig.\,\ref{Fig1} displays the development of the LEED pattern with successive Te deposition and post-annealing to 750\,K after each step (other preparation recipes are discussed in the Appendix. The k-space area around the $(\overline{\nicefrac{1}{3}}~\overline{\nicefrac{1}{3}})$ position is enlarged in the purple enframed inset at the lower left corners. Below each LEED image a corresponding intensity profile taken along the \mbox{[1 1]} k-space direction (cf. red line in Fig.\,\ref{Fig1}(b) is depicted. 

Starting point was the recently discovered and analyzed \threerectcell\ phase \cite{Kisslinger2020}, which is fully developed at a coverage of $\Theta = 0.33$\,ML. 
Fig.\,\ref{Fig1}(a) shows the LEED pattern of this phase, which is a superposition of the contributions of three 120$^\circ$ rotational domains (one reciprocal unit mesh of each is displayed in red, blue, and green). 
The phase is highly ordered as indicated by both the low background and the sharpness of fractional order spots. The latter is not only visualized in the zoom displayed in the inset but even better in the intensity profile below. 
Due to a glide symmetry plane within the structure half-order spots along the principal reciprocal axes like the $(\overline{\nicefrac{1}{2}}~0)$ spot, whose position is marked in yellow in Fig.\,\ref{Fig1}(a), are extinct at all electron energies. An IV-analysis revealed that the surface structure consists of \cutwotetwo\ chains in hcp-positions  of the substrate with a mutual distance of three Cu(111) lattice parameters (for more details see \cite{Kisslinger2020}).

Already for a Te overexposure of less than 0.01\,ML additional sharp spots at tenth order positions appear in the LEED pattern (e.g. the $(\overline{\nicefrac{3}{10}}~\overline{\nicefrac{3}{10}})$ reflex), which are displayed Fig.\,\ref{Fig1}(b) for a higher excess coverage of 0.035\,ML in order to improve their visibility. 
This indicates the coexistence of another equally well-ordered phase with different periodicity. 
This new phase is fully developed at 0.40\,ML Te, where all LEED spots of the former \threerectcell\ phase have completely disappeared as proven by the inset of the LEED pattern as well as the intensity profile of Fig.\,\ref{Fig1}(c).  

The new phase has a rectangular unit cell also, however, with the longer side now increased from 3a$_{Cu}$ to 5a$_{Cu}$, the reciprocal meshes are accordingly smaller as seen from a comparison of Figs.\,\ref{Fig1}(a) and (c). Using the same terminology than for the \threerectcell\ phase, we label the new phase as \rectcell. Note that contrary to the $(3\times \sqrt 3 )_\textrm{rect} \cong (2\sqrt 3 \times \sqrt 3)R30^{\circ}$ the new cell cannot be designated by the Wood notation, only the less convenient matrix notation $\bigl( \begin{smallmatrix} 5 & 0\\	1 & 2 \end{smallmatrix} \bigr) $ would be a proper alternative.
Hence, the size of the unit cell is ten times that of the Cu(111) substrate, i.e. with a coverage of 0.40\,ML we must have four Te atoms per unit cell.

Upon close inspection of the \rectcell\ LEED pattern we find that all half order spots along the principal k-space axes are systematically extinguished for normal incidence of the primary beam, i.e. there is vanishing intensities at those positions for all electron energies. This tells that we have also here -- as for the  \threerectcell\ phase -- a glide symmetry plane along the short basis vector of the real space unit cell ([$\bar1\bar1 2$] direction). 

From experiments with variable annealing temperatures we find indications that the \rectcell\ phase entails long-range mass transport. Even with optimal 0.40\,ML Te coverage a lower annealing temperature than 750\,K leads to streaks in the LEED spot indicating the lack of long-range order in one direction. Below 470\,K the \rectcell\ phase does not develop at all, only a poorly ordered ($2\sqrt 3 \times \sqrt 3$)R30$^\circ$ LEED pattern is observed. For more details see the Appendix.

\subsection{Real-Space Imaging}\label{sec:res:STM}

\begin{figure*}[htb]
	\centering
	\includegraphics[width=0.95\textwidth]{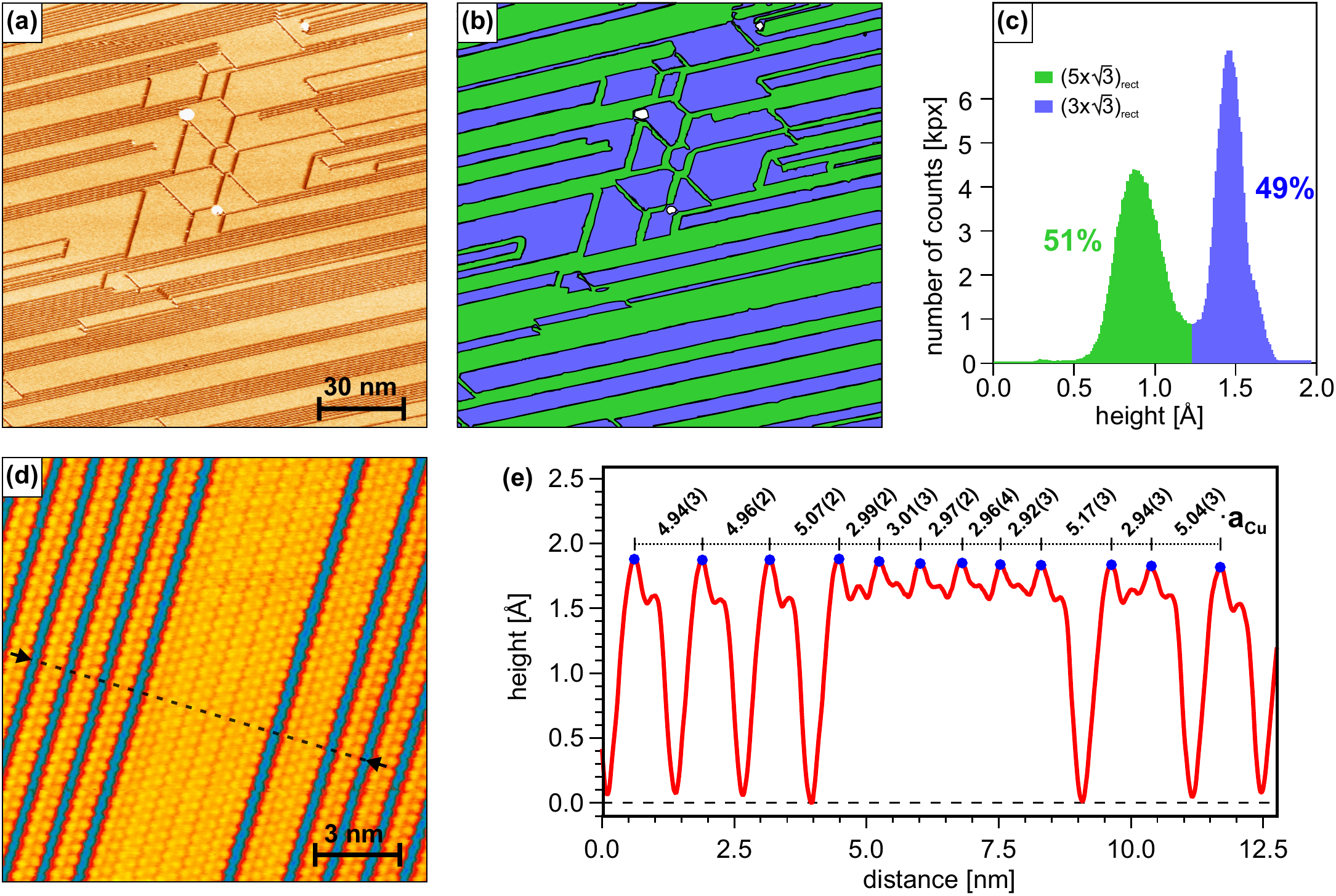}
	\caption{STM imaging for 0.365 ML Te on Cu(111) corresponding to the LEED-pattern of Fig. \ref{Fig1}(b). (a): Large scale (150\,nm)$^2$ image taken at  a flat terrace showing the co-existence between two differently imaged phases. Brighter areas are associated with the \threerectcell\ and darker regions with the \rectcell\ phase.
		(b),(c): Phase area assignment for the image from (a) by means of thresholding and height distribution.
		(d): Atomically resolved (15\textsc\,nm)$^2$ image with phase boundaries. Both structures exhibit zig-zag chains of protrusions as a basic structural element.
		(e): Average of 15 line profiles taken parallel to the black line indicated in (d) through identical atomic positions. All inter-chain distances (given as multiples of the Cu lattice parameter a$_{Cu}$) within and between the two phases are close to either $3a_{Cu}$ and $5a_{Cu}$, errors are obtained via the standard deviation of peak distances within single curves from the mean value. 
		Imaging parameters: (a) $U = 1.7$\,V, $I = 0.50$\,nA; (d) $U = -0.05$\,V, $I = 0.40$\,nA.}
	\label{Fig2}
\end{figure*}

Real-space imaging by means of STM gives access to the local geometrical properties and composition of the surface unit cell and thus reduce the number of physically meaningful models to be tested in a LEED-IV analysis. Furthermore, it gives a much better and even quantitative information on the quality and homogeneity of the preparation of the phases.

\subsubsection{Phase mixture of \threerectcell\ and \rectcell}\label{sec:growth}

First, we will describe the STM results of a surface corresponding to the LEED pattern depicted in Fig.\,\ref{Fig1}(b) with a nominal Te coverage of 0.365 ML, for which we detected the coexistence of both, the \threerectcell\ and \rectcell\ phases.

Large scale STM images on flat terraces ([Fig.\,\ref{Fig2}(a)) reveal two differently imaged regions at the same surface level. Darker appearing areas show stripes with a mutual distance of $5a_{Cu}$ and thus we assign this region to the \rectcell\ phase. Brighter zones occur on that scale as practically structure-less regions and we deduce that they must comprise the \threerectcell\ phase observed in LEED. Via smoothing and thresholding we can easily evaluate the relative areal shares of the two phases as demonstrated in Fig.\,\ref{Fig2}(b),(c). The resulting relative coverage was found to be $49\%$\ and $51\%$\ for the \threerectcell\ and \rectcell\ phase, respectively. This is in perfect agreement with the amount of deposited Te, which was chosen right in the middle between the nominal coverages of 0.33\,ML and 0.40\,ML for the two phases. The correspondence of phase area ratio and average Te coverage also indicates that both phases only exist in a very narrow regime of the phase diagram, i.e. absent or surplus Te atoms must be energetically very expensive in both phases. And indeed, STM does not reveal any vacancy or interstitial Te atom within the phases, cf. Figs.\,\ref{Fig2}(d) and \ref{Fig3}(b).

Within the \threerectcell\ phase regions there are several domain boundaries visible with a mutual angle of 120$^\circ$ according to the threefold symmetry of the Cu(111) substrate. Remarkably, however, is that those 120$^\circ$ boundaries are not observed in between the \rectcell\ phase region and also phase boundaries are typically parallel to the chain direction. This together with the large aspect ratio of \rectcell\ domains let us conclude that non-parallel domain boundaries of this phase must also be energetically very unfavorable. 

Further information about the new \rectcell\ phase comes from a comparison of its local geometrical properties to those of the \threerectcell\ structure, which can be revealed from atomically resolved STM images of two phase regions as shown in Fig.\,\ref{Fig2}(d).

First, it is striking that both structures are imaged quite similar as zig-zag chains oriented along [$\bar1\bar1 2$], which is also the direction of the glide-symmetry plane for both structures. Thus, we conclude that a key element of the new \rectcell\ structure must be the very same \cutwotetwo\ chains known from the \threerectcell\ structure with Te imaged as bright protrusions (cf. \cite{Kisslinger2020}).

A line profile perpendicular to the chains ([$1\bar1 0$]-direction) is plotted in (Fig.\,\ref{Fig1}(e)), representing an average over 15 parallel curves measured through identical atomic positions.
All indicated lateral distances, given as multiples of the lateral lattice parameter a$_{Cu}$ of the substrate, are mean values of differences of corresponding maxima positions within the single curves with their standard deviation taken as statistical error. First we find that the inter-chain separation is always close to either $3a_{Cu}$ or  $5a_{Cu}$. Only at the phase boundaries it appears slightly expanded which we attribute to either different local Te-relaxations in both structures or to a lateral deviation of the center of the imaging orbital to the true atomic position.

 A switch between fcc and hcp position would require a relative lateral shift of $\nicefrac{1}{2}\, a_{Cu}$ along [$1\bar1 0$] and  $\nicefrac{1}{3}\, a_{Cu}$ along [$\bar1\bar1 2$], which would easily be recognized by STM. Thus, we infer that the visible atoms in both phases are located in the same hcp registry w.r.t. the substrate (provided that the chain site is not altered for such narrow domains). Also bridge and top sites can be excluded by similar arguments. They would also lead to significantly different vertical atomic positions in contradiction to our STM observation.

Another important information from the height profile of Fig.\,\ref{Fig1}(e)  is the apparent depth of the troughs between the chains of the \rectcell\ phase. Of course, they strongly depend on tunneling parameters like voltage, tip-sample distance and tip sharpness. Deepest troughs were observed for very low voltages (as used for the STM image in Fig.\,\ref{Fig2}(d)) with depressions as large as about 1.8 \AA. From that we conclude that the area between the chains must lie (at least) one full layer below the chain level.

\subsubsection{Closed \rectcell\ layer}\label{sec:full_layer}

\begin{figure}[htb]
	\centering
	\includegraphics[width=0.55\columnwidth]{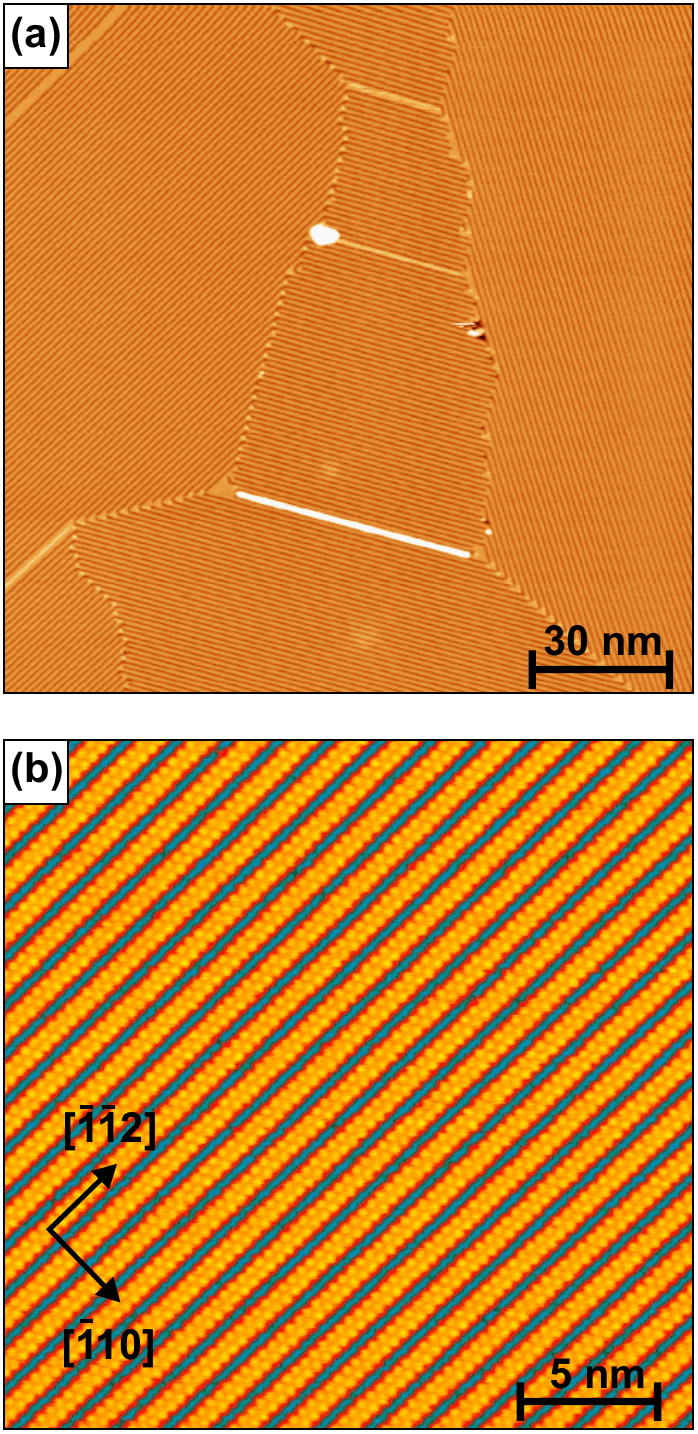}
	\caption{STM appearance of the fully developed \rectcell\ phase for 0.40 ML Te on Cu(111), corresponding to the LEED-pattern depicted in Fig. \ref{Fig1}(c).
		(a): Large scale (150\,nm)$^2$ image taken at a flat terrace presenting all three domains of an otherwise perfectly ordered \rectcell\ phase.
		(b): (25 \,nm)$^2$  atomically resolved image demonstrating that the phase is practically defect free also at the atomic level.
		Imaging parameters: (a) $U = 0.48$\,V, $I = 0.15$\,nA; (b) $U = -0.33$\,V, $I = 0.50$\,nA.}
	\label{Fig3}
\end{figure}

A surface with a Te coverage of 0.40 ML is completely covered by a perfectly ordered \rectcell\ phase, which appears at large scale in three different orientations due to the threefold symmetry of the substrate. This is demonstrated in the wide area STM image of Fig.\,\ref{Fig3}(a), which was recorded at a very extended terrace of the Cu(111) substrate and can be taken as a proof of the high quality and uniformity, with which this \rectcell\ phase can be prepared.

At the atomic level (Fig.\,\ref{Fig3}(b)) we observe the  zig-zag chains oriented along [$\bar1\bar1 2$] and equivalent directions, but even for such perfect samples we were not able to resolve atomic features within the troughs between the \cutwotetwo\ chains despite all our best efforts.
That means, assuming that every bright protrusion in the chains corresponds to a Te atom (as it was for the \threerectcell\ phase), that we ``see'' in STM only two Te atoms per unit cell, which is just half of the deposited amount. 
Apparently, another two Te atoms (referred to as Te$_{\text{hidden}}$) remain hidden somewhere within the unit cell, either in the troughs between or just below the chains themselves. 
This uncertainty allows for a variety of different structural models, among which other methods with better sensitivity to buried layers have to discriminate. 
In section \ref{sec:struct} we will apply both experimental (LEED-IV) and theoretical (DFT) methods to identify the true structural model for the \rectcell\ phase. 

\subsubsection{Exposure beyond $\Theta = 0.40$\,ML}

\begin{figure}[htb]
	\centering
	\includegraphics[width=\columnwidth]{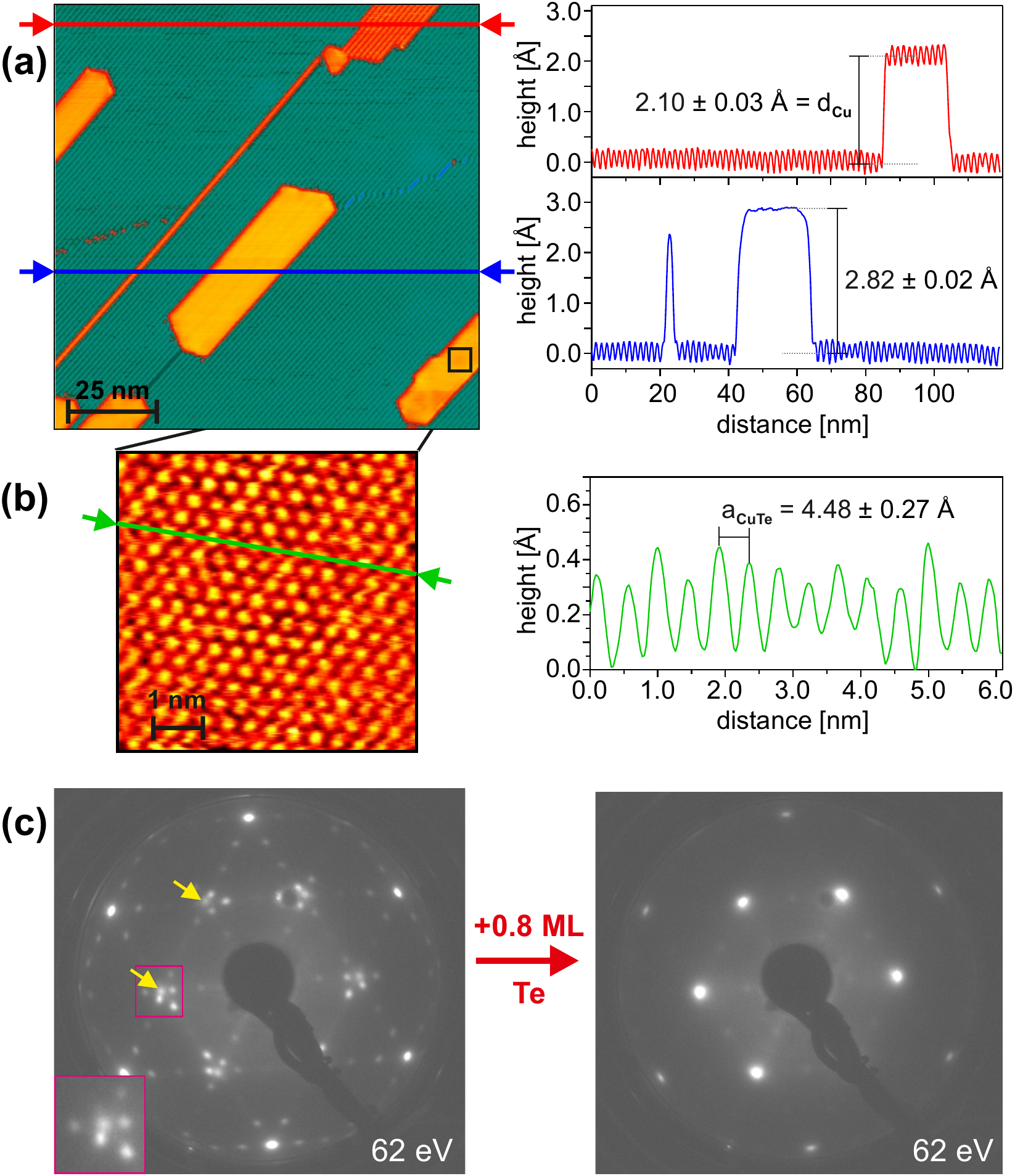}
	\caption{ Data on the phase evolution of Te/Cu(111) for $\Theta > 0.4$\,ML. 
		(a): STM-image and height profiles taken along the red and blue lines for a Te coverage slightly above $\Theta = 0.40~$ML. The dark green areas correspond to the \rectcell phase, the bright areas to a new phase.  ($U = 1.6$\,V, $I = 0.15$\,nA).
		(b): Atomically resolved STM-image for the bottom right  island in (a) and line profile taken along the green line showing a hexagonal array of protrusions with mutual distance of $\approx\sqrt 3 \cdot a_{Cu}$ ($U = 0.42$\,V, $I = 0.50$\,nA).
		(c): LEED-pattern for a Te coverage of $\approx0.7$\,ML (left) showing a mixture of \rectcell\ and a ``$\left(\sqrt{3} \times \sqrt{3}\right)\textrm{R30}^\circ$'' (spots marked by yellow arrows and magnified in the inset) phases. At about 1.5\,ML (left) no residues of the \rectcell\ phase can be found anymore by LEED.}
	\label{Fig4}
\end{figure}

As a final point here, we investigate the consequences of Te overexposure of the \rectcell\ phase. 
From Fig.\,\ref{Fig4}(a) we see that there are islands of a new phase developing in between the still quite perfectly ordered \rectcell\ phase. Their apparent height in STM is about 2.8\,\AA\ above the level of the \rectcell\ phase (blue profile), which can easily be distinguished from other islands, where the \rectcell\ phase is developed on a monolayer high Cu terrace (red profile, apparent height about 2.1\,\AA). Obviously, there is no energetically favorite way to accommodate further Te atoms within the \rectcell\ phase. Hence, its stability regime must be of vanishing width in the phase diagram, enclosed by extended two-phase regimes.
For the island phase, atomically resolved STM images as shown in Fig.\,\ref{Fig4}(b) reveal a hexagonal array of protrusions with a mutual distance of about 4.5\,\AA, which is close to $\sqrt 3 \cdot a_{Cu}$. 
Consequently, additional LEED spots near third order positions appear. 
They are marked by yellow arrows in Fig.\,\ref{Fig4}(c) (left) where the overexposure was chosen higher ($\Theta \approx 0.7$\,ML) for better visibility of the extra spots. The two-phase regime extends to $\Theta \approx 1.5$\,ML as judged from the LEED pattern of  Fig.\,\ref{Fig4}(c) (right).

Guided by the lattice parameter of about 4.5\,\AA\ as well as the apparent thickness of (at least) 5\,\AA\ above the Cu(111) surface we tentatively assign this new phase as a more or less bulk-like copper telluride phase, which will be discussed in more detail in a forthcoming publication.

\section{\label{sec:struct} Crystallographic Structure of the \rectcell\ Phase} 

\subsection{Model discrimination}

\begin{figure}[htb]
	\centering
	\includegraphics[width=\columnwidth]{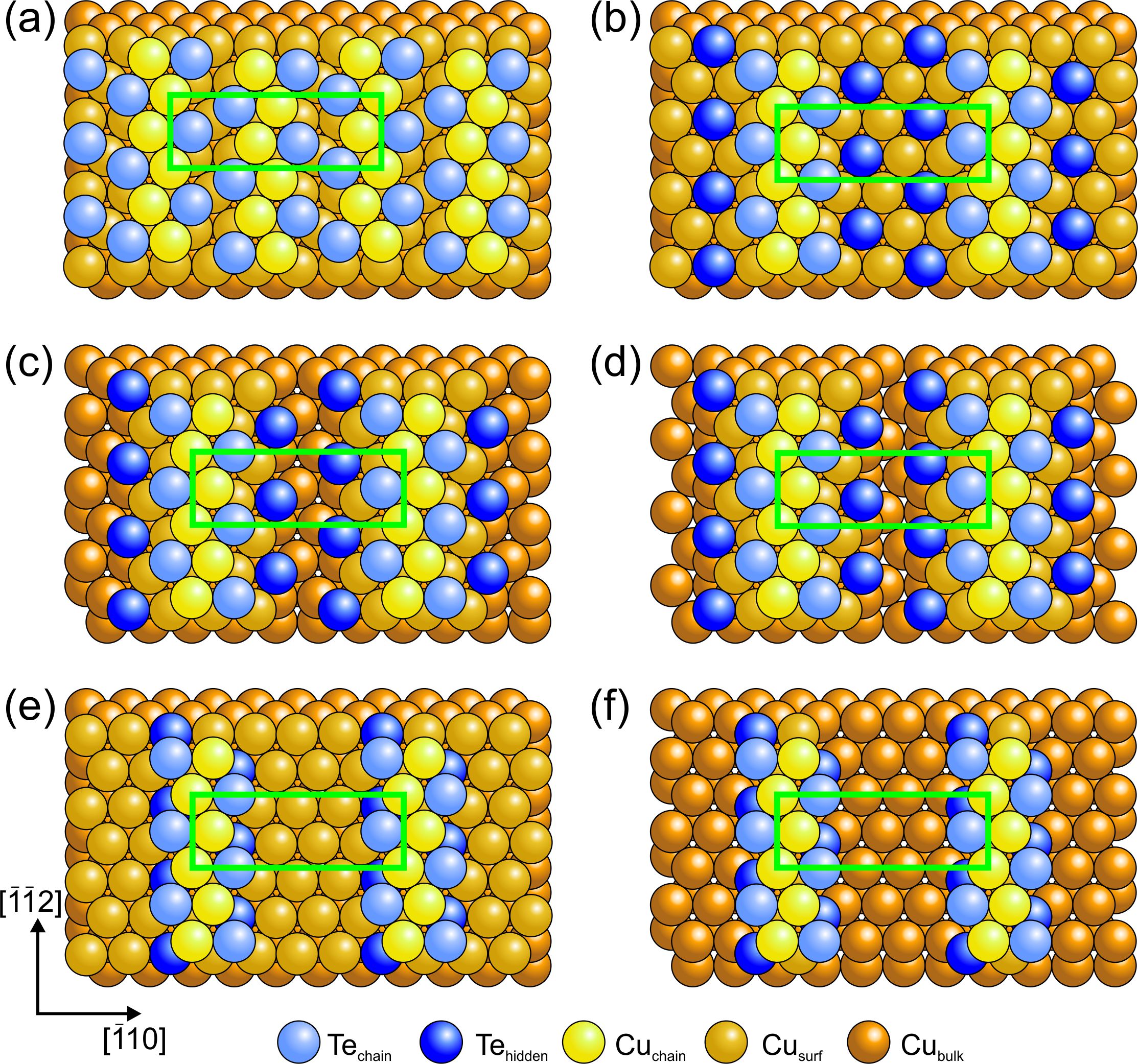}
	\caption{Structural model types tested in LEED and DFT. (a) \cutwotetwo\ chains alternatingly assuming hcp or fcc sites. (b) \cutwotetwo\ chains with extra Te atoms (Te$_{\text{hidden}}$) in substitutional sites in between. (c) Same as (b) but with Cu zig-zag row between Te$_{\text{hidden}}$ atoms removed. (d) Same as (c) but with another Cu zig-zag row from the 2$^{nd}$ Cu layer removed. (e) \cutwotetwo\ bilayer chains embedded into the substrate's top layer. (f) Free-standing \cutwotetwo\ bilayer chains. Structure (c) turns out to be the best-fit structure. For more details see main text.}
	\label{Fig5}
\end{figure}

Visual LEED and STM imaging provides us with two constraints in the search for the structural model of the \rectcell\ phase. 
First, there must be a glide plane along the [$\bar1\bar1 2$] direction linking every two atoms within the unit cell by symmetry.
Second, we identify the same \cutwotetwo\ chains as for the \threerectcell\ phase which at best could exhibit a shift in the stacking sequence (from hcp to fcc).  
The remaining questions are, where the other two Te atoms are located and whether there are any modifications at the underlying Cu(111) substrate. 
For that we tested various structural model types depicted in Fig.\,\ref{Fig5} both by a coarse LEED-IV analysis and via DFT energy minimization as described in Sections\,\ref{sec:leed_calc} and \ref{sec:dft_calc}. 
We elaborate here the steps that lead to the identification of the true surface structure depicted in Fig.\,\ref{Fig5}(c) to emphasize the necessity to combine the information from the three methods employed to reduce the phase space of potential structures considerably and quickly.

Starting from the knowledge of the \threerectcell\ structure one could expect just an alternate sequence of hcp and fcc located \cutwotetwo\ chains as displayed in Fig.\,\ref{Fig5}(a). 
This would allow a denser packing of chains compared to exclusive hcp site occupation as in the \threerectcell\ phase. 
Also energetically such a registry shift of half of the chains should not be very expensive, for the \threerectcell\ this shift was calculated to be only 0.03\,eV per unit cell higher in energy \cite{Kisslinger2020}. 
However, it turned out that the LEED-IV analysis clearly discards this model with an R-factor as high as $R_{(a)} = 0.78$. 
Also, the energy of this phase is 0.74\,eV higher than the structure we identify as the bestfit structure below. Obviously, there is already significant repulsion between the Te atoms of neighboring chains, when their distance is decreased compared to the \threerectcell\ phase. 
Finally, a DFT-based simulation of the STM image predicts both pairs of Te atoms to be equally visible, in clear contrast to experiment (see previous section). 

From STM experiments and the DFT-based STM simulation of that model, one can conclude that the Te$_{\text{hidden}}$ atoms must sit (at least) one layer below the plane of the \cutwotetwo\ chains. 
On the other hand, bulk substitutional sites are energetically highly unfavorable for Te in Cu ($\Delta E = 2.2$\,eV  w.r.t. a surface layer substitutional site). 
Thus, we can restrict our model search with good confidence to sites within the outermost Cu(111) layer. 

\begin{table}[htbp]
	\caption{Compilation of figures of merit for the different tested models. Columns from left to right: Model type corresponding to Fig.\,\ref{Fig5}. Sites assumed by the \cutwotetwo\ chains. R-factors of the LEED Fit for the restricted optimization (c.f. Sec.\ref{sec:leed_calc}). Number $n$ of surplus Cu atoms (with self energy -3.73\,eV) in the model and energetic difference $\Delta E_{f}$ according to Eq.\,\ref{eq:formationE} with respect to the LEED bestfit model ``(c) hcp'' (in bold).}
	\renewcommand{\arraystretch}{1.4}	\setlength{\tabcolsep}{1.5ex}
	\begin{tabular}{cccrc}
		\hline \hline
		Model& Site  & R$_\text{coarse}$ & $n$ & $\Delta E_{f}$[eV] \\	
		\hline 
		(a)  &         & 0.78 & +6 & 0.74  \\
		(b)  & hcp     & 0.55 & +2 & 0.67  \\
		& fcc     & 0.80 & +2 & 0.69  \\
		\textbf{(c)}   & \textbf{hcp} & \textbf{0.29} & \textbf{0} & \textbf{ 0} \\			
		& fcc     & 0.77 &  0 & 0.04  \\
		(d)  & hcp     & 0.44 & -2 & 1.06  \\
		& fcc     & 0.78 & -2 & 1.04  \\
		(e)  & hcp     & 0.66 & +2 &   --- \\
		& fcc     & 0.84 & +2 & 3.44  \\
		(f)  & hcp/hcp & 0.78  & +6 & 2.38  \\
		\hline \hline
	\end{tabular} 
	\label{dftleed}
\end{table}

With these guidelines the most obvious sites for the Te$_{\text{hidden}}$ atoms are substitutional ones right between the chains as shown in Fig.\,\ref{Fig5}(b) with a mutual Te-Te distance of $\sqrt 3 \cdot a_{Cu}$, just as within the \cutwotetwo\ chains. 
However, the still very high R-factor value $R_{(b)} = 0.55$ obtained suggests that the model as such cannot be correct. 
Intriguingly, the LEED-IV fit wants to place the Te$_{\text{hidden}}$) atoms to a much lower position (by about 0.8\,\AA) than predicted by DFT, i.e. almost at the level of the Cu surface layer. 
This, however, appears un-physical because of the larger size of Te atoms compared to Cu atoms. The only way to make room for them is to remove the neighboring Cu atoms. This leads to the model displayed in Fig.\,\ref{Fig5}(c). Here, the Te atoms can shift a bit aside in order to assume proper bond distances to both the neighbored Cu atoms and those of the layer below. The LEED-IV fit for this structure improves a lot to $R_{(c)} = 0.29$ and also the DFT calculation finds a formation energy (Eq.\,\ref{eq:formationE}) being 0.67\,eV (fcc: 0.69\,eV) lower than for model (b), cf. Tab.\,\ref{dftleed}. 

Since removing Cu atoms for making room for the Te$_{\text{hidden}}$ atoms turned out to improve the fit we further tested a structure in which also one zig-zag row of Cu atoms from the second substrate layer is removed (Fig.\,\ref{Fig5}(d)). 
This model, however, is disfavored by the LEED analysis with $R_{(d)} = 0.44$ and also by DFT with an increase of the formation energy by 1.06\,eV (fcc: 1.04\,eV).

For all models (b) -- (d) the respective fcc variants are almost degenerate with respect to their DFT energies, but exhibit much worse R-factor values in the LEED analyses, cf. Tab.\,\ref{dftleed}. They can therefore be discarded just on the basis of this restricted LEED geometry optimization. 

An alternative class of models can be constructed when the Te$_{\text{hidden}}$ atoms are placed right below the chains forming some kind of bilayer chain as shown in Figs.\,\ref{Fig5}(e),(f).
Of course,  with these it would be hard to understand the strict 5-fold periodicity of the chain arrangement. 
Those bilayer chains can either be embedded into the Cu(111) top layer (model (f)) or free-standing (model (g)). 
It turns out during the DFT relaxation that the two chains strongly separate from each other in vertical direction, which would hardly be a geometrically stable configuration. 
This is also expressed by the extraordinarily high formation energy of these models (3.44\,eV and 2.38\,eV for models (e) and (f) after 100 relaxation steps, cf. Tab.\,\ref{dftleed}). Therefore, we also could forego to test all possible combinations of fcc and hcp stackings between the chains and towards the substrate. In line with the DFT results the LEED fits also led to very high R-factor values between 0.66 -- 0.84, again ruling out this class of models.

 Based on the variance of the R-factor, which is due to the enormous data base as low as var(R) = 0.012 for the analysis of the hcp variant of model (c), we can rule out any other of the tested models with highest confidence. This is in line with the fact that DFT finds this structure as the by far lowest in energy. Even more, for this model the fitted structural parameters of the LEED-IV analysis did not much deviate from the values of the relaxed DFT structure (see Section \ref{sec:structure}) in contrast to any other tested model. Since there are hardly any other conceivable structural models, we are confident that with  model (c) we have indeed found the true ground state structure for 0.40\,ML of Te on Cu(111). The model exhibits \cutwotetwo\ ad-chains on hcp sites, which are already known from the preceding \threerectcell\ phase. Additionally, there is a severe surface reconstruction, whereby four Cu atoms per unit cell are expelled from the surface layer, forming wide troughs, which accommodate two Te atoms per unit cell. We will therefore refer to this model as \underline{A}d-\underline{C}hain-and-\underline{T}rough (ACT) model. Different to most of the other models, the DFT-based STM image simulation reveals a perfect agreement with experiment as shown in Fig.\,\ref{Fig6}(a)

\subsection{\label{sec:structure}Detailed Structure of the ACT-Phase}

Fitting the LEED-IV spectra with a finer parameter grid and including further lateral, vibrational and non-structural parameters we finally succeeded to reduce the R-factor level for the ACT-model to a value of $R_{ACT} = 0.174$  ($0.077/0.187$ for integer/fractional order beams). This value still appears quite high regarding the perfectness, with which the phase can be prepared. For comparison, the match of experimental and computed IV-spectra was  with $R = 0.099$ \cite{Kisslinger2020} much better for the similarly well prepared \threerectcell\ phase of Te on this substrate. 

\begin{figure}[tb]
	\centering
	\includegraphics[width=0.98\columnwidth]{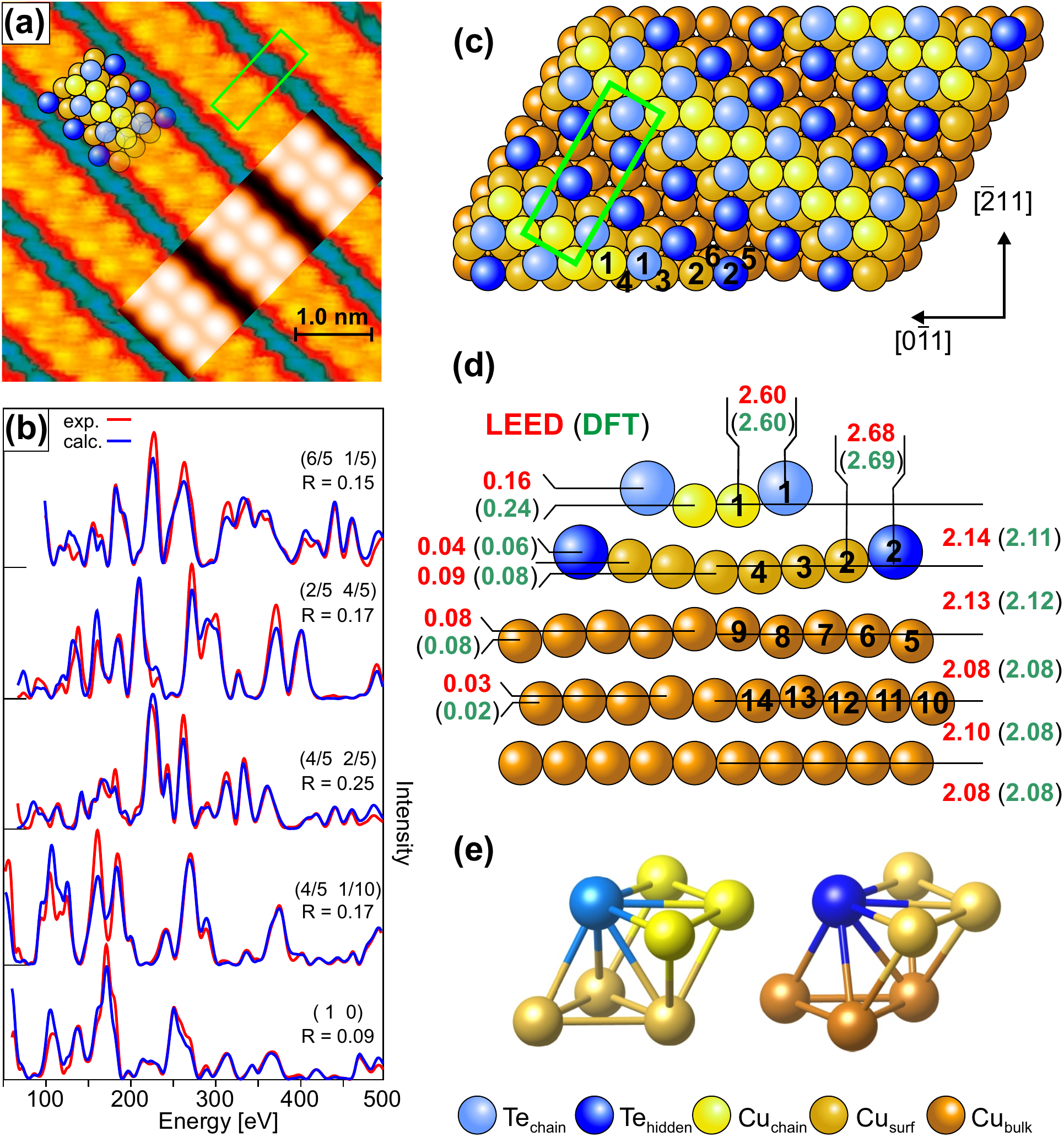}
	\caption{Results of the structure analysis for the \rectcell\ phase of Te on Cu(111).
		(a) Experimental STM image ($U = -0.33$\,V, $I = 0.50$\,nA) partly overlaid by a DFT-based STM simulation and a ball model. (b) Selection of best-fit LEED IV-spectra. (c,d) Ball model of the bestfit structure: (c) top view, (d) side view along the [$\overline 2$11] direction (vertical distances are strongly exaggerated). Red (green) numbers denote average Cu layer spacings, Te-Cu vertical bucklings and bond lengths (all in \AA) as derived from LEED-IV (DFT). (e) Visualization of the unusual local binding geometry of Te atoms.}
	\label{Fig6}
\end{figure}

A reason for this insufficiency might lie in the LEED-IV calculation scheme, where both the real and imaginary part of the inner potential are taken spatially constant below a boundary parallel to the surface. For a structure as rough as in the present case, however, this is a quite insufficient description of the true wave propagation. 
As a worst-case example, an electron being scattered from one \cutwotetwo\ chain to the next, will virtually propagated through vacuum over a distance of about 6\,\AA. Within this section, its kinetic energy (and wave length) is determined by the vacuum level and there is also hardly any damping. In the LEED calculation, in contrast, the whole path will be treated as being inside the crystal, where the kinetic energy and the attenuation are altered by the real and imaginary part of the inner potential. For this particular wave path this leads to an erroneous phase shift of $0.72\pi$/$0.12\pi$ and an attenuation factor of 0.47/0.77 at 50/500\,eV kinetic energy, respectively. Of course, other scattering paths will be much less affected by this approximation. Nevertheless, it is understandable that no perfect agreement between experimental and calculated LEED-IV spectra can be achieved for this particular structure especially in the lower energy range. This is visualized in Fig.\,\ref{Fig6}(b) for a selection of beams with single beam R-factor values close to the total R-factor. The full set of 79 beams is given in the Supplemental \mbox{Material \cite{SupMat}}. 

Analyzing the structure in detail we find a 0.16\,\AA\ buckling within the \cutwotetwo\ chains, which is due to the larger size of the Te atoms compared to Cu atoms, in analogy to the \threerectcell\ phase \cite{Kisslinger2020}. 
At variance to that phase we now have the next Te atoms not within the adjacent chain but somewhat closer (and at an asymmetric position) within the troughs, cf. Fig.\,\ref{Fig6}(c). 
Regarding the various local atomic shifts, we find clear indication for a significant repulsion between these different types of Te atoms. At first, the whole chain is shifted along the [$\bar1 \bar1 2$] direction by about 0.2\,\AA\ in order to equalize the Te\,1 -- Te\,2 distances as far as possible. 
Second, the upper Te\,1 atom is pressed against the adjacent Cu chain atoms (bond lengths 2.60-2.66\,\AA\ compared to 2.63-2.71\,\AA\ for the \threerectcell\ phase), while the lower lying Te\,2 is pushed away towards the center of the trough with the consequence that the whole Cu$_6$Te$_2$ stripe of the second layer is widened (Cu-Cu distances increased by 0.02-0.10\,\AA\ w.r.t. bulk value).

The chemical impact of the Te adatoms as well as the trough reconstruction of the Cu(111) substrate also induces a remarkable buckling not only within the Cu stripe right below the \cutwotetwo\ chains (0.09\,\AA) but also within the two subsequent complete and close-packed layers below (0.08\,\AA\ and 0.03\,\AA). 
The average layer distances, however, remain almost bulk-like, c.f. Fig.\,\ref{Fig6}(d). The detailed vertical and lateral displacements of all varied atomic positions are listed in the Supplemental Material\,\cite {SupMat} together with the respective error margins derived from single parameter ``error curves''. 

Comparing the experimentally derived coordinates with those predicted by DFT (scaled by the ratio of experimental and theoretical lattice parameters), we find a very good correspondence except for the coordinates of the Te atoms and in parts of adjacent Cu\,1 atoms, cf. Fig.\,\ref{Fig6}(d) and tabulated values in the Supplemental Material\,\cite{SupMat}. This was similarly observed in case of the \threerectcell\ phase and is an indication that in DFT the Te atoms are somewhat larger than in reality and by that overestimating the buckling amplitudes. Moreover, it appears that the DFT also underestimates the Te-Te repulsion and therefore ends up with a smaller \cutwotetwo\ chain shift and a lesser second layer Cu stripe expansion. These discrepancies might serve as a standard for future optimization of the DFT functionals for Te atoms.

\section{\label{sec:discussion} Discussion}

Our work unravels the formation and structure of the new ACT phase with \rectcell\ structure on Cu(111) with 0.40 ML Te content. 
Due to the high (post-) annealing temperatures (470\,K\,$<$\,T\,$<$\,750\,K) needed for preparation of the phase, it might have been overlooked so far in various studies dealing with submonolayer amounts of Te on Cu(111) \cite{Andersson1968, Comin1982, King2012, Lahti2014, Tong2020}.
The question remains what drives the formation of this \rectcell\ phase. For that, we first analyze the bonding scheme of the Te atoms. The formation of the troughs within the Cu(111) surface leads to [210]-oriented mini-facets. All Te atoms both in ad-chains as well as within the troughs form six bonds towards Cu, cf. Fig.\,\ref{Fig6}(e). Three of these bonds are parallel to the (111)-plane with bond lengths of 2.60...2.66\,\AA\ (ad-chain) and 2.60...2.77\,\AA\ (trough) and three are out-of-plane towards the Cu layer below (2.66...2.82\,\AA\ for ad-chains  and 2.59...2.78\,\AA\ in the troughs). The only difference between upper and lower Te atoms is that the local binding clusters are mirrored at the drawing plane of Fig.\,\ref{Fig6}(e), which comes from the fcc$\rightarrow$hcp registry shift of the ad-chains. Disregarding local relaxations of Cu atoms the Te atoms assume semi-crystal locations (ad-sites with highest possible coordination) which can be regarded either as a [$\bar{1}\bar{1} 1$] step site at the (111) surface or, alternatively, as a kink site of a stepped (100) plane. 

Obviously, this particular and seemingly energetically most favorable bonding configuration requires the existence of local atomic steps. 
For the \threerectcell\ phase this was accomplished via the formation of the \cutwotetwo\ chains, where the Cu core just resembles such a [$\bar{1}\bar{1} 1$] step. 
However, this mechanism comes to an end at a Te coverage of 0.33\,ML just for steric reasons. 
Only by trough formation a somewhat higher amount of Te  can be accommodated, since the registry shift between subsequent fcc layers shortens the lateral distance of respective adsorption sites. 
The creation of such channels within the Cu(111) terraces requires a rather high activation energy and thus high preparation temperatures, since expelling atoms from an intact close-packed (111) terrace is energetically very costly. 
In contrast, for the \cutwotetwo\ chain formation it is sufficient to capture diffusing Cu adatoms, which have to be thermally detached from step edges. Incidentally, the high energetic prize to be paid for the creation of the steps also requires that these sites have to be completely saturated by Te atoms. That is why this \rectcell\ phase can only exist right at the nominal Te coverage of 0.4\,ML as found in Sec.\,\ref{sec:full_layer}. It is also an explanation for the finding that 120$^\circ$ domain boundaries (c.f. Sec.\,\ref{sec:growth}) are very rare, since the crossing of troughs would produce quite a couple of odd sites.

One could ask then, why this facetting of the surface does not continue for increasing Te coverage. In principle, this scheme can be continued by increasing the trough width successively by $2 a_{Cu}$ and the depth accordingly by one more layer, forming $\left( (2n+3) \times \sqrt3 \right)_{rect}$ (with $n \in \mathbb{N}_0$) structures. For large $n$, this would eventually lead to a Te coverage of $\Theta = 0.5$\;ML. 
However, the realization fails not only due to the enormous activation energy required for the removal of that many Cu atoms in order to build up extended facets at the surface. Also the average Te site energy will increase in such a configuration: firstly, an increasing share of Te atoms have to assume  the less favorable fcc sites, since only the terminating \cutwotetwo\ chain can remain at the energetically favorable hcp position. 
Otherwise, wide Cu stripes would have to shift to hcp sites as well, which appears highly unlikely. 
Secondly, also no lateral relaxation of Te and Cu atoms, which was observed for the \rectcell\ phase in order to relieve the Te-Te repulsion, would be possible at extended facets. As a result, the \rectcell\ phase turns out to be the most dense surface telluride phase achievable on Cu(111). Exceeding this coverage limit, precipitates of a 3D phase start to grow.  

As a final point, we want to mention that Zhou et al. \cite{Zhou2020}, performing co-adsorption experiments of Sb and Te on Cu(111), also observed chain structures with the very same chain distances of either $3a_{Cu}$ or $5a_{Cu}$ and striking STM resemblance to the ones observed here. 
However, with only STM and DFT results at hand, they interpreted the observed structure as pure Te atomic wires detached from the substrate (Te-Cu bond length $\geq 3\,\textrm{\AA}$). 
This is presumably one more example that the correct structure of a surface phase cannot be safely determined on the mere basis of a combined STM and DFT analysis without a sound crystallographic analysis of the system. Of course, any further prediction of physical or chemical properties of such a system (e.g. electronic properties of atomic Te chains) will collapse, when based on a false structural model.

\section{\label{sec:summary} Summary}

We presented a comprehensive picture of the formation process and the energetics of a so far unknown submonolayer phase of copper telluride with \rectcell\ periodicity, which evolves on Cu(111) for $\Theta = 0.40~$ML. We performed a complete quantitative LEED-IV analysis of this phase that resulted in close parameter accordance with the predictions of a corresponding DFT calculation. By that we obtain detailed and unequivocal knowledge of the structure of this phase. 
It is characterized by \cutwotetwo\ chains combined with a displacive surface reconstruction, whereby 4-atom-wide troughs are formed hosting  two of the four Te atoms per unit mesh (ad-chain-and-trough model). Finally, we prove and rationalize that this new \rectcell\ phase represents the most dense surface telluride phase possible on Cu(111) prior to the onset of bulk telluride formation.

\begin{acknowledgments}
We gratefully acknowledge supply of computing resources and support provided by the Erlangen Regional Computing Center (RRZE). We also cordially thank Prof. Michael Schmid, TU Vienna, for his support in extracting the LEED-IV spectra from the LEED videos.
\end{acknowledgments}

\vspace{1cm}

\appendix*

\section{Preparation Recipes for the \rectcell\ Phase}

\begin{figure*}[htb]
	\centering
	\includegraphics[width=0.95\textwidth]{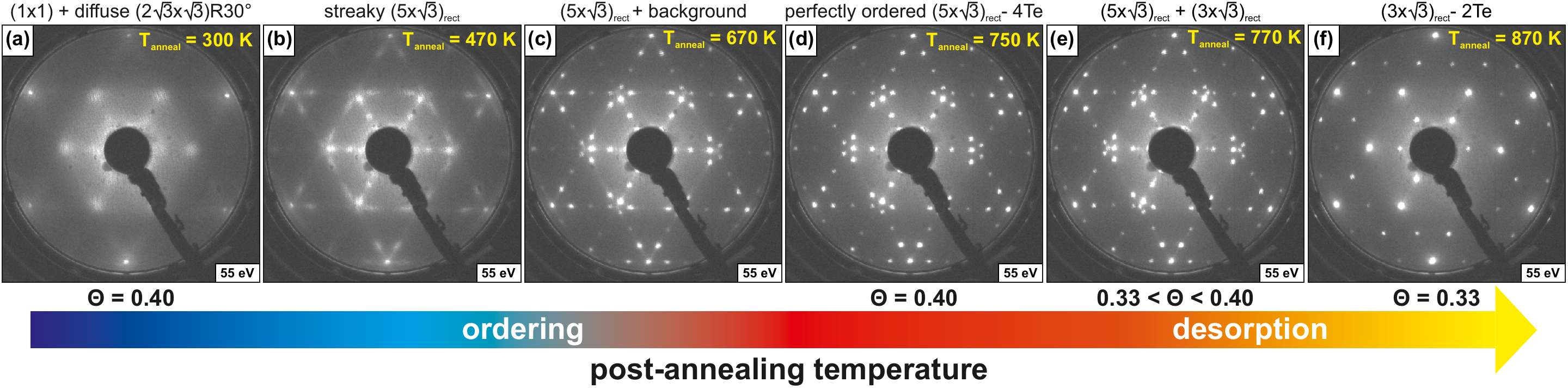}
	\caption{Series of diffraction images for $\Theta = 0.40~$ML Te deposited on Cu(111) at 90\,K and successively annealed to the different indicated temperatures for 1\,min each.  LEED images were taken at an electron energy of $55~$eV and the sample cooled back to 90\,K.}
	\label{Fig7}
\end{figure*}

In Sec.\,\ref{sec:res:leed} it was shown that the \rectcell\ phase can be prepared from the well-ordered \threerectcell\ phase via deposition of additional 0.07\,ML Te and annealing to 750\,K. However, there is no need to prepare the \threerectcell\ phase first, which will be dissolved anyway. One can just start with the total amount of 0.40\,ML Te deposited at the cold Cu(111) sample. Right after deposition at 90\,K there is hardly any order to be observed. Post-annealing to room temperature (or direct deposition at this temperature) leads to a diffuse ($2 \sqrt3 \times \sqrt3$)R30$^\circ$ LEED pattern displayed in Fig\,\ref{Fig7}(a). Obviously, room temperature is not sufficient to overcome the barrier for expelling substrate atoms from a plain Cu(111) terrace. For that it needs an annealing temperature of about 470\,K, where a weakly ordered \rectcell\ phase develops first (Fig\,\ref{Fig7}(b)). With rising annealing temperature the degree of order also increases gradually (Fig\,\ref{Fig7}(c)), presumably because of the emerging long-range Cu mass transport, until perfect order is generated at 750\,K (Fig\,\ref{Fig7}(d)). For only slightly higher temperatures like 770\,K or for prolonged annealing just at 750\,K the onset of Te desorption is observed leading into the  regime of phase mixture with the \threerectcell\ phase (Fig\,\ref{Fig7}(e)). Eventually, after annealing to 870\,K the \rectcell\ phase has completely vanished (Fig\,\ref{Fig7}(f)).

It should be noted that the \rectcell\ phase can also be prepared from Te over-exposure via direct heating to 750\,K, whereby all excess Te desorbs. However, such a procedure is quite tricky since the \rectcell\ structure itself is not stable against further Te desorption at those temperatures as outlined before. Thus it is delicate to choose the right breakpoint of the annealing process.

\bibliography{literatur_TeCu111-5xsqrt3_rect}

\end{document}


\preprint{APS/123-QED}

\title{Supplemental Material:\\ New submonolayer copper telluride phase on Cu(111) - ad-chain and trough formation}

\author{Tilman Ki{\ss}linger}
\author{M. Alexander Schneider}%
\email{alexander.schneider@fau.de}
\author{Lutz Hammer}%
\email{lutz,hammer@fau.de}
\affiliation{%
	Solid State Physics, Friedrich-Alexander-Universität Erlangen-Nürnberg, Staudtstraße 7, 91058 Erlangen, Germany
}%

\date{\today}

\begin{abstract}
	\vspace{1.5cm}

	This supplement contains tabulated parameter values of the LEED bestfit with errors and corresponding DFT values. Further, the full set of experimental LEED intensity data in comparison with their calculated bestfit counterparts is presented in order to visualize the quality of correspondence.  Additionally, ``error curves''  for all parameters adjusted in the course of the fit procedure are provided. Here, each parameter is varied around its bestfit value with all other kept fix. The parameter range, where the R-factor value remains below the variance level $R+var(R)$, is taken as range of uncertainty \cite{Pendry1980}. As separate files are attached a cif-File for visualisation of the bestfit structure and the experimental IV-data set formatted for input in the new VipErLEED package \cite{Riva2021}, whereby the normalized beam intensities are aligned row-wise separated by semicolons.\\
	
\end{abstract}

\maketitle

\bibliography{literatur_TeCu111-5xsqrt3_rect_SM}

\clearpage

\section{\label{LEED_struct} Details of the Crystallographic Structure}

Table\,\ref{Bestfit_values} summarizes all structural and non-structural parameters adjusted in the course of the LEED-IV analysis together with their theoretical counterparts predicted by the DFT calculation. For better clarity we give the deviations of atomic coordinates from respective bulk lattice positions (for the \cutwotetwo\ chain atoms the lateral positions are given w.r.t. the ideal hcp site). Additionally, the table also comprises the vibrational amplitudes fitted independently for the two inequivalent pairs of Te atoms as well as for under-coordinated Cu atoms (Cu\,1,2,5). The bestfit values for the inner potential were V$_{0i} = 3.80\,\text{eV}~^{\text{+0.61\,\text{eV}}}_{\text{$-$0.65\,\text{eV}}}$ and V$_{00} = 2.50\,\text{eV}~^{\text{+0.42\,\text{eV}}}_{\text{$-$0.44\,\text{eV}}}$ and the effective angle of incidence was adjusted to $\Theta_\text{eff} = 0.35^\circ~^{\text{+0.16}^\circ}_{\text{$-$0.33}^\circ}$.

Apart from the positions of Te atoms and directly neighbored Cu atoms (see discussion in the main paper) experimental and theoretical values coincide much closer than the error margins determined from the variance of the R-factor. This corroborates the more general finding that this error estimate is usually quite conservative. In particular, when many parameters enter the fit it would be a quite improbable scenario, when the whole error margin would be used up by one single parameter with all others remaining right at their bestfit value. The close correspondence between LEED and DFT parameters can also be taken as a proof that the fit of that many parameters can safely be performed when using such a huge data base as in the present case. As already discussed in the main paper, the deviation of determined Te positions from the DFT prediction is most probably owed by an overestimation of the ``size'' of Te atoms in DFT, which also affects the equilibrium position of attached Cu atoms.

\begin{table*}[b]
	\caption {Structural parameters (in \AA) resulting from the LEED IV-analysis for surface atoms defined in Fig.\,\ref{Fig6} are gives as deviations $\Delta x$, $\Delta y$, and $\Delta z$ from the respective bulk positions (for Te\,1 and Cu\,1 the lateral positions are w.r.t. the hcp site). 		
		Corresponding values predicted by DFT scaled by the ratio of experimental and calculated lattice parameters (factor 0.9923) are given in the neighboring columns. Note that only one out of each pair of symmetry-linked atoms is mentioned here. For the experimental parameters the errors margins were derived from the intersection of single parameter R-factor curves with the variance level (R+var(R)), see Section \ref{errors}. Vibrational amplitudes $\Delta x$ (in \AA), where fitted, are given in the last column.} 
	\renewcommand{\arraystretch}{1.4}	\setlength{\tabcolsep}{1.5ex}

	\vspace{3mm}
	
	\begin{tabular}{c@{\hspace{8mm}}rr@{\hspace{8mm}}rr@{\hspace{8mm}}rr@{\hspace{10mm}}r}
		\hline \hline
		{atom} & \multicolumn{2}{c}{$\Delta x$~~} & \multicolumn{2}{c}{$\Delta y$} & \multicolumn{2}{c}{$\Delta z$}& \multicolumn{1}{c}{$\Delta u$}\\[.2ex]
		& \multicolumn{1}{c}{LEED}& \multicolumn{1}{c}{DFT~~~~} & \multicolumn{1}{c}{LEED}& \multicolumn{1}{c}{DFT~~~~}  & \multicolumn{1}{c}{LEED}& \multicolumn{1}{c}{DFT~~~~}& \multicolumn{1}{c}{LEED~~} \\[.5ex]
		\hline                                                                                                  	
		Te\,1        & 0.108~$^{\text{+0.034}}_{\text{$-$0.036}}$    &  0.100 & 0.204~$^{\text{+0.051}}_{\text{$-$0.043}}$   & 0.107& 0.260~$^{\text{+0.015}}_{\text{$-$0.015}}$  & 0.301 & 0.115~$^{\text{+0.023}}_{\text{$-$0.017}}$  \vspace{1.5mm} 
		\\                                                                                        
		Te\,2        &0.281~$^{\text{+0.044}}_{\text{$-$0.036}}$    &0.243 & $-$0.014~$^{\text{+0.044}}_{\text{$-$0.039}}$   & $-$0.044 & 0.120~$^{\text{+0.014}}_{\text{$-$0.014}}$  &0.125 & 0.085~$^{\text{+0.023}}_{\text{$-$0.036}}$  \vspace{1.5mm} 
		\\                                                                                             
		Cu\,1        & $-$0.003~$^{\text{+0.043}}_{\text{$-$0.036}}$    &  $-$0.006 & 0.187~$^{\text{+0.042}}_{\text{$-$0.040}}$  & 0.089 & 0.102~$^{\text{+0.013}}_{\text{$-$0.010}}$  & 0.057 & 0.129~$^{\text{+0.011}}_{\text{$-$0.011}}$ \vspace{1.5mm} 
		\\                                                                                             
		Cu\,2        &0.109~$^{\text{+0.043}}_{\text{$-$0.041}}$    &  0.028 & 0.034~$^{\text{+0.039}}_{\text{$-$0.062}}$   & $-$0.013 & 0.082~$^{\text{+0.022}}_{\text{$-$0.013}}$ 	& 0.065 & ~~~0.125~$^{\text{+0.019}}_{\text{$-$0.017}}$ \vspace{1.5mm}
		\\
		
		Cu\,3        & 0.062~$^{\text{+0.038}}_{\text{$-$0.033}}$    & 0.031 & 0.033~$^{\text{+0.057}}_{\text{$-$0.066}}$   & 0.003& 0.053~$^{\text{+0.021}}_{\text{$-$0.014}}$  & 0.030   &   \vspace{1.5mm} 
		\\
		
		Cu\,4        & 0.014~$^{\text{+0.038}}_{\text{$-$0.045}}$    &  0.008 & 0.024~$^{\text{+0.049}}_{\text{$-$0.046}}$   & 0.002 & ~$-$0.008~$^{\text{+0.015}}_{\text{$-$0.013}}$  & $-$0.011   & \vspace{1.5mm} 
		\\
		
		Cu\,5        & 0.041~$^{\text{+0.042}}_{\text{$-$0.040}}$    &  0.023& $-$0.030~$^{\text{+0.047}}_{\text{$-$0.045}}$    & $-$0.026& $-$0.044~$^{\text{+0.016}}_{\text{$-$0.016}}$  & $-$0.051 & 0.091~$^{\text{+0.021}}_{\text{$-$0.016}}$ \vspace{1.5mm} 
		\\
		
		Cu\,6        & 0.004~$^{\text{+0.039}}_{\text{$-$0.040}}$    &  $-$0.010 & $-$0.002~$^{\text{+0.059}}_{\text{$-$0.065}}$   & $-$0.012 & $-$0.016~$^{\text{+0.024}}_{\text{$-$0.022}}$  & $-$0.006 & \vspace{1.5mm}
		\\
		
		Cu\,7        & $-$0.015~$^{\text{+0.047}}_{\text{$-$0.046}}$    &  $-$0.019 & $-$0.009~$^{\text{+0.050}}_{\text{$-$0.054}}$  & $-$0.026 & 0.021~$^{\text{+0.014}}_{\text{$-$0.019}}$  & 0.015 & \vspace{1.5mm}
		\\
		
		Cu\,8        & $-$0.002~$^{\text{+0.029}}_{\text{$-$0.035}}$   	&  $-$0.008  & 0.017~$^{\text{+0.066}}_{\text{$-$0.051}}$   & $-$0.009 & $-$0.024~$^{\text{+0.019}}_{\text{$-$0.015}}$  & $-$0.013 & \vspace{1.5mm}
		\\
		
		Cu\,9        & 0.001~$^{\text{+0.039}}_{\text{$-$0.035}}$    &  0.001 & 0.026~$^{\text{+0.044}}_{\text{$-$0.044}}$  & $-$0.001 & 0.036~$^{\text{+0.012}}_{\text{$-$0.016}}$  & 0.029 & \vspace{1.5mm} 
		\\
		
		Cu\,10        & 0.017~$^{\text{+0.053}}_{\text{$-$0.060}}$    &  0.003 & 0.005~$^{\text{+0.051}}_{\text{$-$0.059}}$   & 0.005& ~$-$0.012~$^{\text{+0.017}}_{\text{$-$0.022}}$  & $-$0.010 & \vspace{1.5mm} 
		\\
		
		Cu\,11       & $-$0.002~$^{\text{+0.052}}_{\text{$-$0.052}}$    &  $-$0.008 & $-$0.020~$^{\text{+0.068}}_{\text{$-$0.069}}$   & $-$0.020& 0.002~$^{\text{+0.028}}_{\text{$-$0.025}}$   & 0.003 & \vspace{1.5mm} 
		\\
		
		Cu\,12       & $-$0.004~$^{\text{+0.057}}_{\text{$-$0.054}}$    &  $-$0.006 & $-$0.016~$^{\text{+0.062}}_{\text{$-$0.065}}$  & $-$0.015 & $-$~0.012~$^{\text{+0.028}}_{\text{$-$0.027}}$   & $-$0.009 & \vspace{1.5mm}  
		\\
		
		Cu\,13       & 0.006~$^{\text{+0.068}}_{\text{$-$0.070}}$    &  $-$0.001 & 0.009~$^{\text{+0.114}}_{\text{$-$0.095}}$   & $-$0.004 & 0.014~$^{\text{+0.068}}_{\text{$-$0.059}}$  & 0.014 & \vspace{1.5mm} 
		\\
		
		Cu\,14       & $-$0.002~$^{\text{+0.078}}_{\text{$-$0.072}}$    &  $-$0.008 & $-$0.032~$^{\text{+0.105}}_{\text{$-$0.062}}$   & $-$0.011& 0.000~$^{\text{+0.060}}_{\text{$-$0.065}}$   & 0.007 & \vspace{1.5mm} 
		\\
		\hline \hline
	\end{tabular}
	\label{Bestfit_values}
	
\end{table*}

\clearpage

\section{Bestfit spectra}

\begin{figure*}[htb]
	\includegraphics[width=0.98\textwidth]{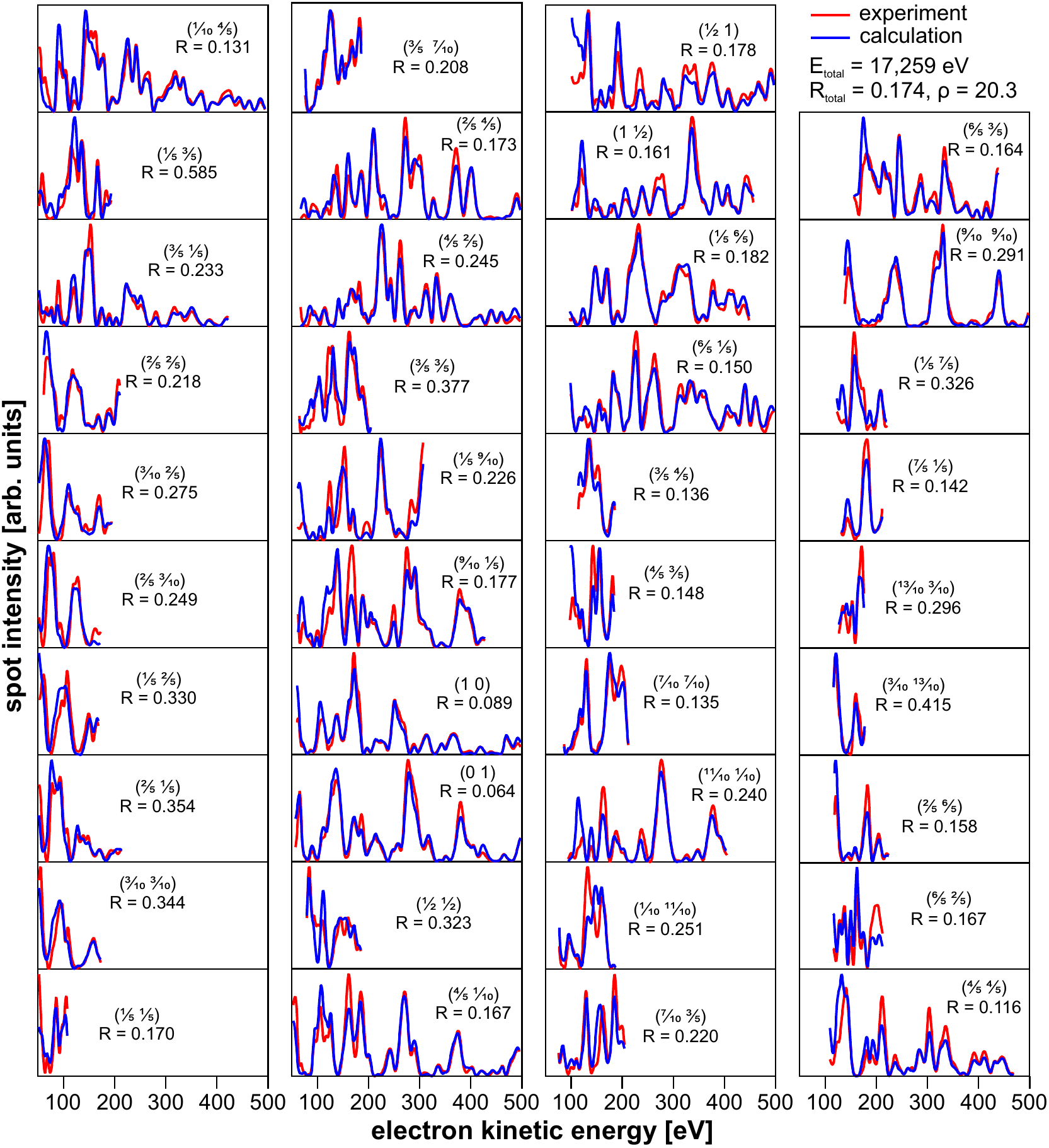}
	\caption{Compilation of all experimental LEED-IV spectra (red) entering the structural analysis together with their calculated counterparts (blue). For better visibility all spectra were normalized to the same average intensity level.} \label{Bestfit_spectra}
\end{figure*}

\newpage

	\includegraphics[width=0.98\textwidth]{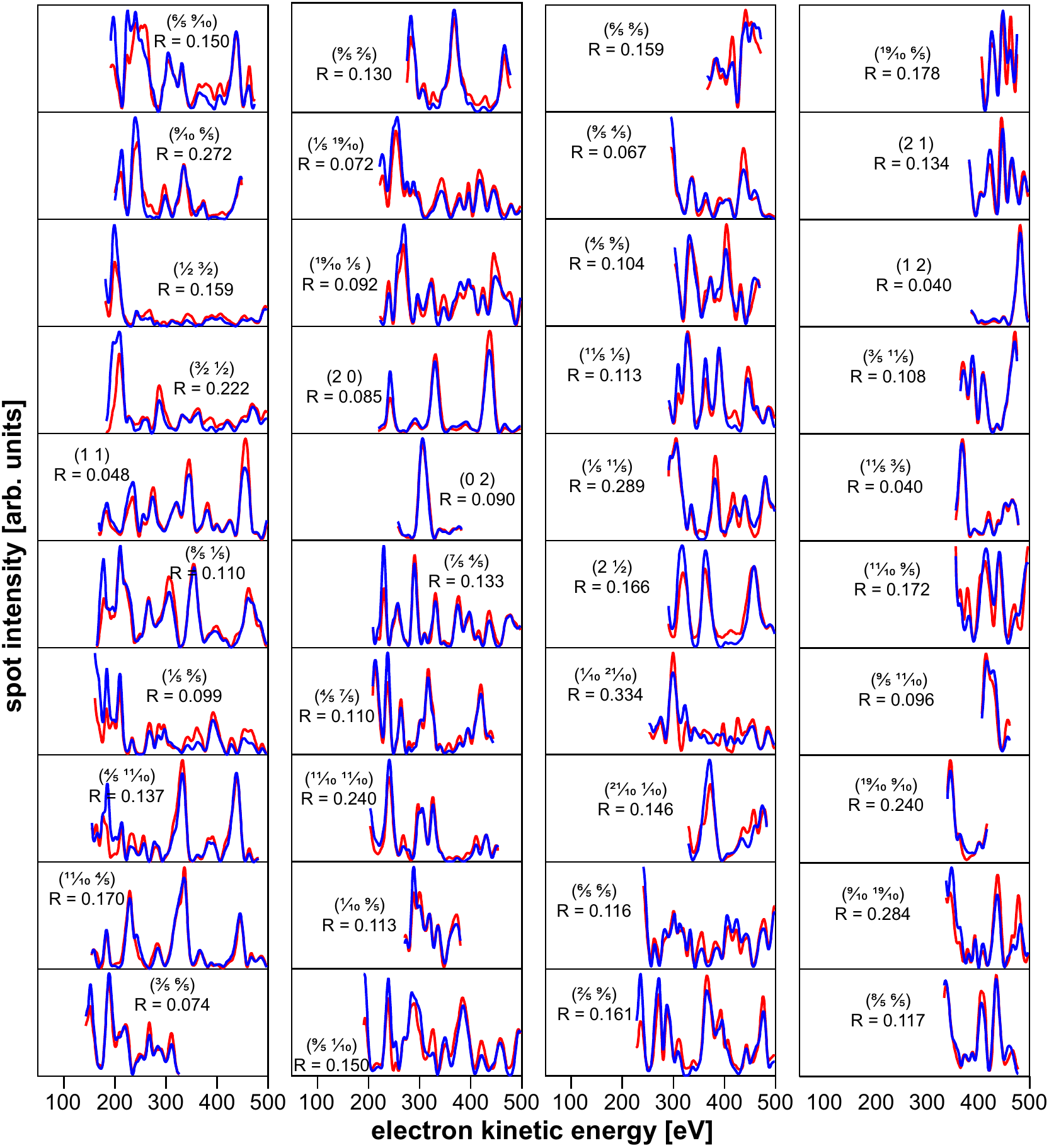}
	
\vspace{5mm}
	FIG. 1 {(continued)}

\newpage

\section{\label{errors}Error curves}

\begin{figure*}[htb]
		\includegraphics[width=0.98\textwidth]{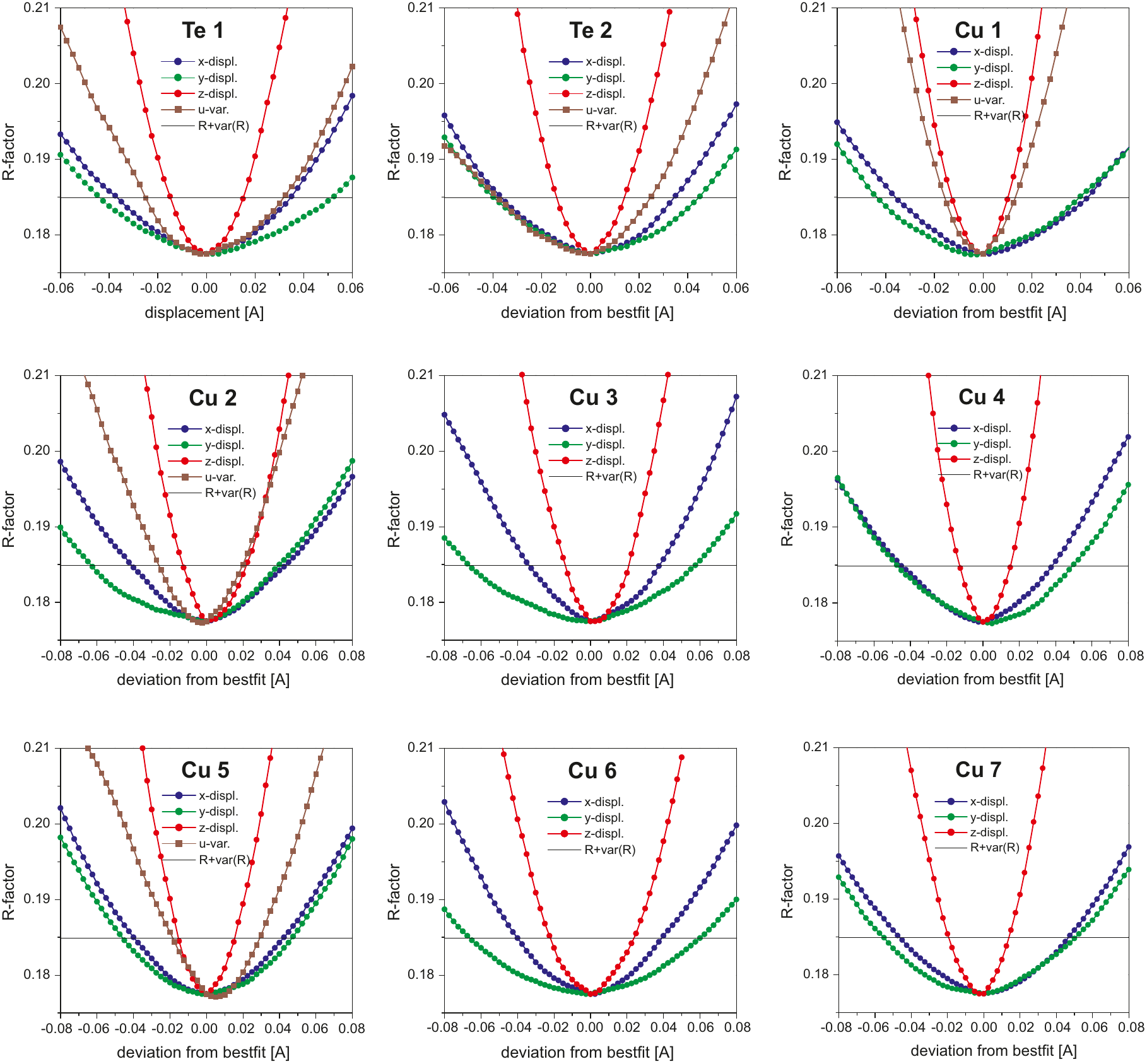}
		\caption{ \noindent R-factor variation of all parameters adjusted in the course of the LEED-IV analysis. For reasons of computational time saving the error analysis was not performed for azimuthally averaged spectra but for one single azimuth of incidence only ($\Phi=180^\circ$).
			Therefore, the error curves as well as the variance level are vertically shifted by $\Delta R = 0.003$ compared to the bestfit.}
		\label{Error_curves}
\end{figure*}

\newpage

		\includegraphics[width=0.98\textwidth]{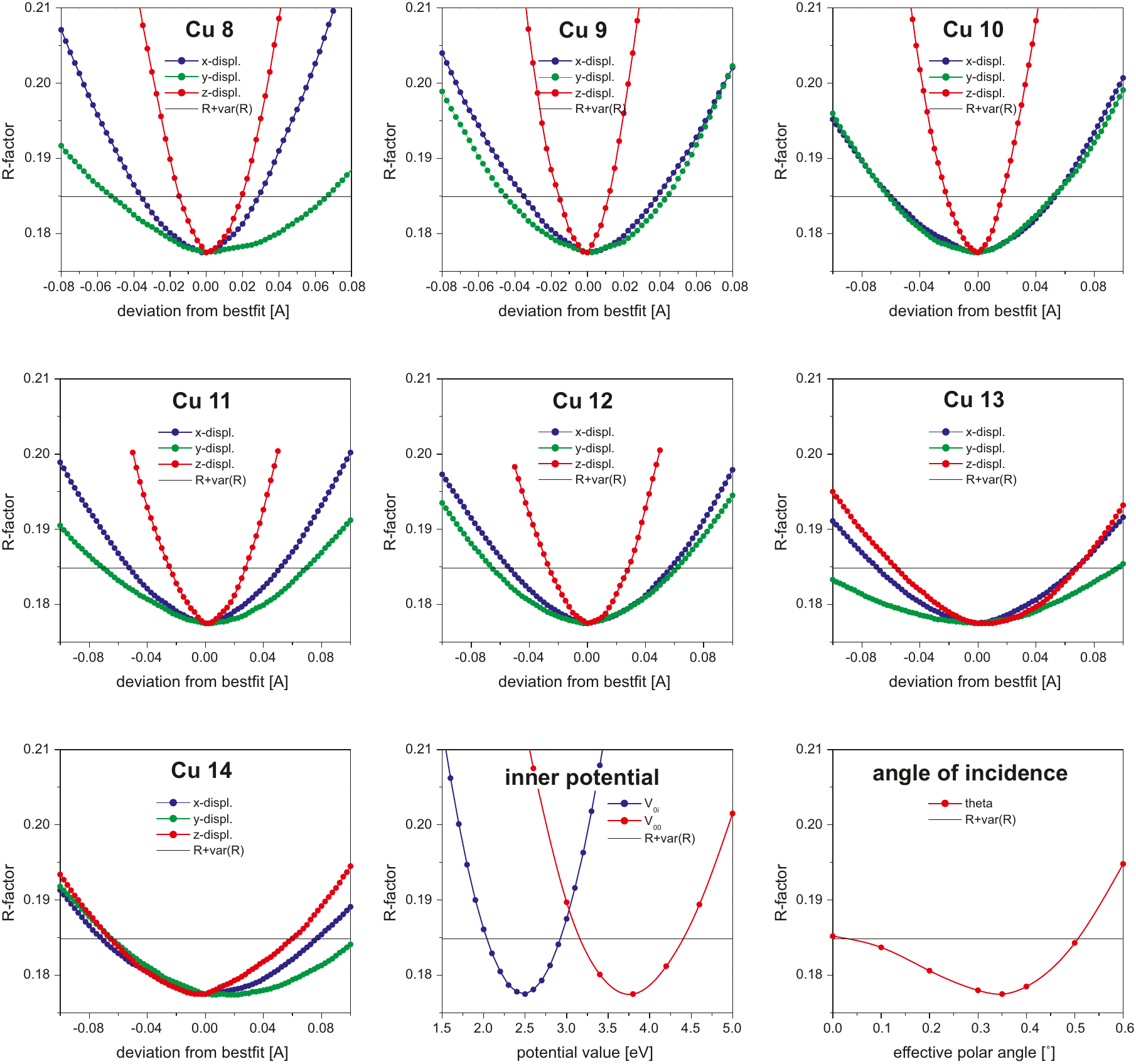}
	
\vspace{5mm}
FIG. 2 {(continued)}